\documentclass[aps,prb,reprint,superscriptaddress,citeautoscript,floatfix,longbibliography]{revtex4-2}
\usepackage{times,siunitx,amsmath,amssymb,amsfonts,hyperref,graphicx,mhchem,orcidlink}
\usepackage[utf8]{inputenc}
\usepackage[T1]{fontenc}
\usepackage[capitalize]{cleveref}
\begin{document}

\title{
Superconductivity induced by structural reorganization\\ 
in the electron-doped cuprate Nd$_{2-x}$Ce$_x$CuO$_4$ 
}

\author{Anita Guarino\orcidlink{0000-0001-8191-854X}}
\thanks{These authors contributed equally to this work}
\affiliation{Dipartimento di Fisica ``E. R. Caianiello'', Università degli Studi di Salerno, 84084 Fisciano (Salerno), Italy}
\affiliation{Consiglio Nazionale delle Ricerche CNR-SPIN, UOS Salerno, 84084 Fisciano (Salerno), Italy}
\author{Carmine Autieri\orcidlink{0000-0002-5008-8165}}
\thanks{These authors contributed equally to this work}
\affiliation{International Research Centre Magtop, Institute of Physics, Polish Academy of Sciences, Aleja Lotników 32/46, 02668 Warsaw, Poland}
\affiliation{Consiglio Nazionale delle Ricerche CNR-SPIN, UOS Salerno, 84084 Fisciano (Salerno), Italy}
\author{Pasquale Marra\orcidlink{0000-0002-9545-3314}}
\email{pmarra@ms-u.tokyo.ac.jp}
\affiliation{Graduate School of Mathematical Sciences, The University of Tokyo, 3-8-1 Komaba, Meguro, Tokyo, 153-8914, Japan}
\affiliation{Department of Physics, and Research and Education Center for Natural Sciences, Keio University, 4-1-1 Hiyoshi, Yokohama, Kanagawa, 223-8521, Japan}
\author{Antonio Leo\orcidlink{0000-0002-9137-2111}}
\affiliation{Dipartimento di Fisica ``E. R. Caianiello'', Università degli Studi di Salerno, 84084 Fisciano (Salerno), Italy}
\affiliation{Consiglio Nazionale delle Ricerche CNR-SPIN, UOS Salerno, 84084 Fisciano (Salerno), Italy}
\affiliation{NANO\_MATES Research Center, Università degli Studi di Salerno, 84084 Fisciano (Salerno), Italy}
\author{Gaia Grimaldi\orcidlink{0000-0001-5438-8379}}
\affiliation{Consiglio Nazionale delle Ricerche CNR-SPIN, UOS Salerno, 84084 Fisciano (Salerno), Italy}
\affiliation{Dipartimento di Fisica ``E. R. Caianiello'', Università degli Studi di Salerno, 84084 Fisciano (Salerno), Italy}
\author{Adolfo Avella\orcidlink{0000-0002-3874-2216}}
\email{a.avella@unisa.it}
\affiliation{Dipartimento di Fisica ``E. R. Caianiello'', Università degli Studi di Salerno, 84084 Fisciano (Salerno), Italy}
\affiliation{Consiglio Nazionale delle Ricerche CNR-SPIN, UOS Salerno, 84084 Fisciano (Salerno), Italy}
\affiliation{Unità CNISM di Salerno, Università degli Studi di Salerno, 84084 Fisciano (Salerno), Italy}
\author{Angela Nigro\orcidlink{0000-0001-8326-5781}}
%\email{nigro@sa.infn.it}
\affiliation{Dipartimento di Fisica ``E. R. Caianiello'', Università degli Studi di Salerno, 84084 Fisciano (Salerno), Italy}
\affiliation{Consiglio Nazionale delle Ricerche CNR-SPIN, UOS Salerno, 84084 Fisciano (Salerno), Italy}

%\date{\today}

\begin{abstract}
Electron-doped and hole-doped superconducting cuprates exhibit a symmetric phase diagram as a function of doping. 
This symmetry is however only approximate.
Indeed, electron-doped cuprates become superconductors only after a specific annealing process:
This annealing affects the oxygen content by only a tiny amount, but has a dramatic impact on the electronic properties of the sample. 
Here we report the occurrence of superconductivity in oxygen-deficient \ce{Nd_{2-x}Ce_xCuO4} thin films grown in an oxygen-free environment, after annealing in pure argon flow.
As verified by x-ray diffraction, annealing induces an increase of the interlayer distance between \ce{CuO2} planes in the crystal structure.
Since this distance is correlated to the concentration of oxygens in apical positions, and since oxygen content cannot substantially increase during annealing, our experiments indicate that the superconducting phase transition has to be ascribed to a migration of oxygen ions to apical positions during annealing. 
Moreover, as we confirm via first-principles density functional theory calculations, the changes in the structural and transport properties of the films can be theoretically described by a specific redistribution of the existing oxygen ions at apical positions with respect to \ce{CuO2} planes, which remodulates the electronic band structure and suppresses the antiferromagnetic order, allowing the emergence of hole superconductivity.
\end{abstract}

\maketitle

%La +3
%Sr +2
%Cu +1 +2 +3
%O 6 (-2)
%Nd +3 +4
%Ce +3 +4

% INTRODUCTION
\section{Introduction}

Since the discovery of superconductivity in LaBaCuO by Bednorz and M\"uller in 1986~\cite{bednorz_possible_1986}, the family of high-temperature cuprate superconductors has grown to include more than hundreds of compounds~\cite{chu_hole-doped_2015} with temperatures as high as \SI{133}{K} at atmospheric pressure~\cite{schilling_superconductivity_1993}.
%However, despite the great theoretical effort, a microscopic understanding of the superconducting pairing mechanism is still lacking. 
These compounds share a similar crystal structure made up of stacked layers of copper-oxygen planes and fit into a universal phase diagram, where superconductivity emerges on doping an antiferromagnetic Mott insulator~\cite{lee_doping_2006,tohyama_recent_2012}.
Indeed, by doping the stoichiometric parent compound via ionic substitution, the antiferromagnetic phase is suppressed, and superconductivity appears.
Ionic substitution may result in the creation of additional holes or electrons in the \ce{CuO2} planes.
Hole-doped~\cite{bednorz_possible_1986,schilling_superconductivity_1993,chu_hole-doped_2015} (e.g., \ce{La_{2-x}Sr_xCuO4}) and electron-doped~\cite{takagi_superconductivity_1989,tokura_electron_1989,armitage_progress_2010,fournier_t_2015,greene_strange_2020} (e.g., \ce{Nd_{2-x}Ce_xCuO4}) share a similar temperature-doping phase diagram, which indicates a common origin of the superconducting pairing. 
However, the symmetry between hole- and electron-doped cuprates is only approximate.
For example, superconductivity in electron-doped cuprates is much harder to achieve since the antiferromagnetic phase persists at higher doping levels~\cite{armitage_progress_2010,fournier_t_2015,horio_suppression_2016,song_electron_2017,helm_correlation_2015}.

Perhaps the most puzzling anomaly of electron-doped cuprates is the fact that doping alone does not produce superconductivity~\cite{armitage_progress_2010,fournier_t_2015,naito_reassessment_2016}.
As-grown samples are antiferromagnetic Mott insulators and become superconducting only after high-temperature oxygen-reducing annealing~\cite{takagi_superconductivity_1989,tokura_electron_1989}.
Annealing reduces the oxygen content by a small fraction~\cite{moran_extra_1989,tarascon_growth_1989,radaelli_evidence_1994,schultz_single-crystal_1996,klamut_properties_1997,navarro_oxygen_2001} (between 0.1\% and 2\%), which decreases the interlayer distance~\cite{kwei_structure_1989,maiser_pulsed_1998,tsukada_role_2005,matsumoto_synthesis_2009,matsumoto_generic_2009,krockenberger_emerging_2013} and contributes to additional electrons in \ce{CuO2} layers~\cite{wei_electron_2016,horio_suppression_2016,song_electron_2017,horio_angle_2018,horio_electronic_2018,lin_extended_2020,ishii_post_2020}.
This results in a dramatic change of the electronic properties~\cite{gauthier_different_2007,richard_competition_2007,dagan_hole_2007,charpentier_antiferromagnetic_2010,asano_ce_2018,li_hole_2019,hirsch_understanding_2019}, including the emergence of the superconducting transition and a reduction of the Néel temperature~\cite{matsuda_magnetic_1992,mang_spin_2004,horio_suppression_2016}, which cannot be achieved only by doping (e.g., adding extra cerium in \ce{Nd_{2-x}Ce_xCuO4}~\cite{matsumoto_synthesis_2009}).
Furthermore, single crystals of the undoped parent compound \ce{Nd2CuO4} are never superconducting.
Conversely, \ce{Nd2CuO4} thin films exhibit superconductivity after annealing, even without doping~\cite{matsumoto_synthesis_2009,naito_reassessment_2016}.
In all cases, the annealing process must be carried in rather specific conditions that drive the samples almost to the limit of decomposition~\cite{kim_phase_1993,mang_phase_2004}.
For these reasons, it is clear that the annealing process must have additional effects. 
These may be the consequence of a reorganization of the crystal structure and/or a change of the distribution of dislocations and defects in the sample, such as the removal of the interstitial apical oxygens (defects)~\cite{radaelli_evidence_1994,xu_oxygen_1996,torrance_role_1989,ohta_apex_1991,higgins_role_2006}, the removal of intrinsic in-plane oxygens~\cite{riou_infrared_2001,riou_pr_2004,richard_role_2004}, or the migration of copper ions to repair and reduce copper vacancies~\cite{kurahashi_heat_2002,kang_microscopic_2007}.
A measurable effect of annealing is the change of the $c$-axis lattice parameter, which is 2 times the interlayer distance between \ce{CuO2} planes:
The lattice parameter decreases to an optimal value $c_\text{SC}$ at which superconductivity appears~\cite{krockenberger_grown_2015,naito_reassessment_2016}.
Generally, oxygen reduction produces a decrease of the $c$-axis parameter associated with the removal of apical oxygen~\cite{radaelli_evidence_1994,schultz_single-crystal_1996}:
Hence, the value of $c$ is considered as a qualitative measure of the oxygen content~\cite{tsukada_role_2005,matsumoto_generic_2009,krockenberger_emerging_2013}.

In this work, we report superconductivity in oxygen-deficient \ce{Nd_{2-x}Ce_xCuO4} (NCCO) thin films obtained by annealing in oxygen-free atmosphere, and we provide a theoretical framework to describe the electronic properties and the structural changes before and after annealing.
Our samples are grown by DC sputter deposition in oxygen-free atmosphere, and exhibit a $c$-axis parameter shorter than the optimal value $c_\text{SC}$, which indicates oxygen deficiency and the presence of a negligible amount of apical oxygens.
Remarkably, these samples become superconducting after annealing in pure argon atmosphere, with a simultaneous \emph{increase} of the $c$-axis parameter.
This strongly indicates that the superconducting phase transition cannot be ascribed to a change of the oxygen content, but a microscopic reorganization of the crystal structure induced by annealing.
Moreover, to obtain a complete phase diagram as a function of the $c$-axis parameter, we have grown thin films also in oxygen/argon atmosphere.
These samples exhibit a $c$-axis parameter longer than the optimal value $c_\text{SC}$ and, as expected, become superconducting after annealing, with a \emph{decrease} of the $c$-axis, in agreement with previous studies~\cite{takagi_superconductivity_1989,tokura_electron_1989,armitage_progress_2010,fournier_t_2015}.
In all samples, the superconductivity appears only when the $c$-axis parameter reaches the optimal value $c_\text{SC} =\SI{12.08}{\angstrom}$.
As we show using first-principles density functional theory (DFT), the evolution of the $c$-axis parameter and the presence of holes can be explained in terms of a microscopic structural modification, i.e., with existing oxygens ions partially migrating to apical positions with respect to \ce{CuO2} planes.
This induces a remodulation of the energy bands and the suppression of antiferromagnetic order, allowing the emergence of hole superconductivity, i.e., the pairing of hole carriers within the same electronic band~\cite{dagan_hole_2007,li_hole_2019,hirsch_understanding_2019}.

\section{Fabrication and characterization}

% CRYSTAL STRUCTURE 

\begin{figure}
\includegraphics[width=1\columnwidth]{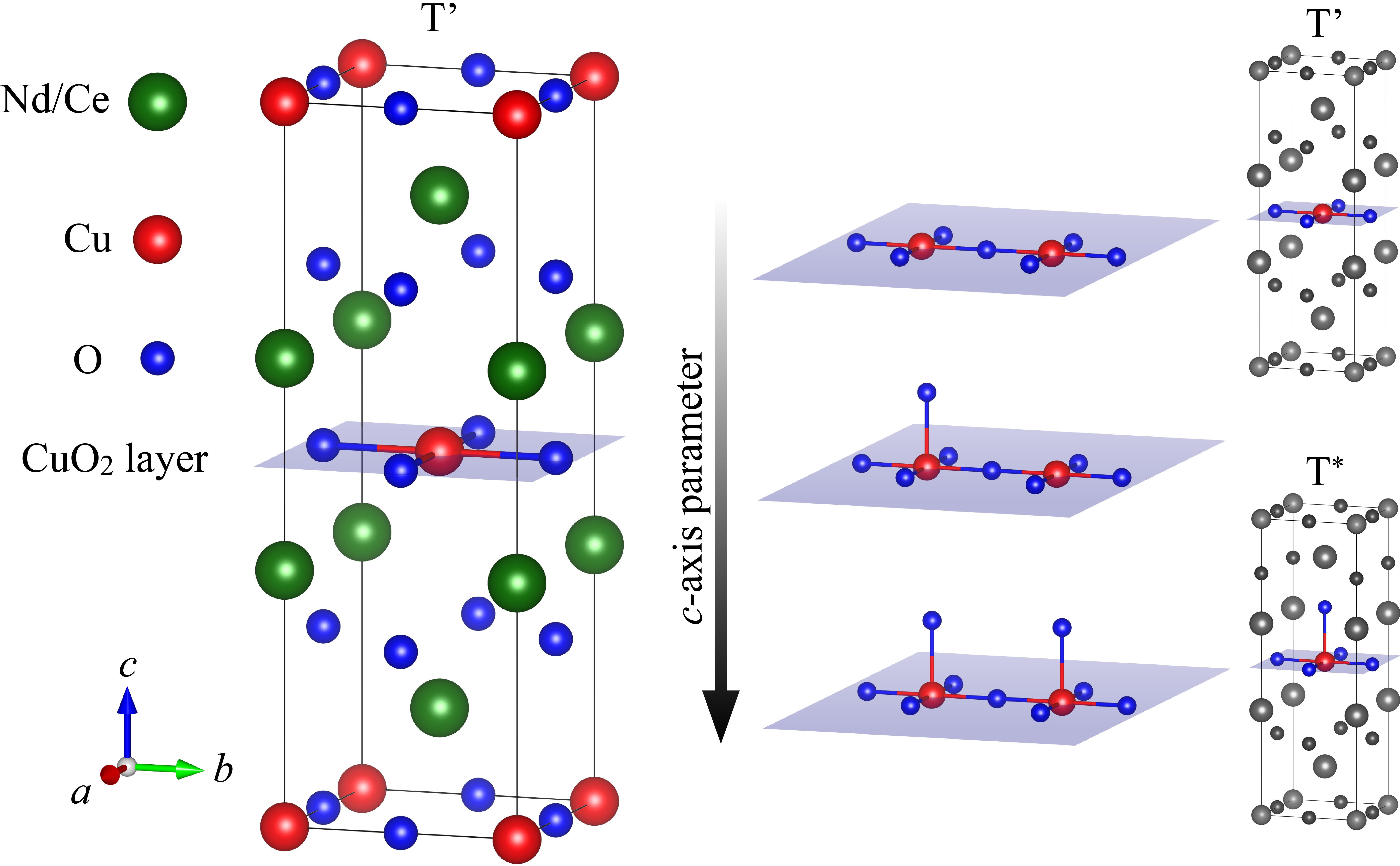}%
\caption{
The NCCO parent compound (undoped \ce{Nd2CuO4}) crystallizes in the tetragonal $\mathrm{T}'$ structure, with
\ce{CuO2} layers sandwiched between charge reservoir \ce{Nd_{2-x}Ce_xO2} layers,
copper ions surrounded by square-planar arrangements of oxygen ions in the $ab$ plane, and
oxygen ions located within the \ce{CuO2} and charge reservoir layers.
In doped compounds, oxygen ions may partially occupy apical sites above or below the Cu atoms, e.g., one apical oxygen for every two copper atoms, or every copper atom ($\mathrm{T}^*$ structure). 
The concentration of apical oxygens is correlated to the elongation of the $c$-axis parameter.
} 
\label{fig:structure}
\end{figure}

The undoped parent compound \ce{Nd2CuO4} crystallizes in a tetragonal $\mathrm{T}'$ crystal structure, containing \ce{CuO2} layers stacked along the $c$-axis and sandwiched between the charge reservoir layers [see \cref{fig:structure}(a)].
Moreover, thin films of NCCO and other electron-doped typically exhibit disorder, with oxygen vacancies (in \ce{CuO2} layers or charge reservoir layers) and excess oxygen at apical sites (above and below \ce{CuO2} layers)~\cite{armitage_progress_2010,fournier_t_2015,naito_reassessment_2016} [see \cref{fig:structure}(b)].
In particular, the presence of in-plane oxygen vacancies is correlated to an increase of electrons in the conduction band~\cite{wei_electron_2016,horio_suppression_2016,song_electron_2017,horio_angle_2018,horio_electronic_2018,ishii_post_2020}, whereas the concentration of oxygen ions on apical sites is correlated with the elongation of the $c$-axis parameter~\cite{tsukada_role_2005,matsumoto_generic_2009,krockenberger_emerging_2013}.

% FABRICATION

Our experiment used optimized DC sputtering to grow well-oriented NCCO films without spurious phases and with a fixed cerium content $x=0.17\pm0.01$. 
We obtained films with thickness $\SIrange{100}{200}{nm}$ grown respectively in pure argon (type A samples) and mixed argon/oxygen atmosphere with \ce{O2}$/$\ce{Ar}$>2\%$ (type B samples), at $\SI{1.7}{mbar}$ total pressure and heater temperature $\SI{850}{\celsius}$ (see also Refs.~\onlinecite{guarino_fabrication_2013,Guarino_2014}). 
After initial \emph{in situ} annealing, we performed a high-temperature \emph{ex situ} annealing at \SIrange{900}{950}{\celsius} for \numrange{0.5}{2} hours, depending on the film thickness.
We deliberately fabricated samples with different growing and annealing conditions to study the interplay between structural reorganization and superconductivity, regardless of other factors (see Appendix).

% XRAY

We measure the lattice parameters by x-ray diffraction (XRD) before and after high-temperature annealing.
We found $c=\SIrange{12.04}{12.07}{\angstrom}$ and $a=b\approx\SIrange{3.95}{3.96}{\angstrom}$ for as-grown type A samples, while $c=\SIrange{12.09}{12.15}{\angstrom}$ and $a=b\approx\SIrange{3.94}{3.97}{\angstrom}$ for as-grown type B samples. 
After annealing, type A samples grown in oxygen-free atmosphere exhibit a slight elongation of the $c$-axis, whereas the in-plane lattice parameter remains unchanged. 
Conversely, type B samples grown in oxygen atmosphere exhibit a systematic decrease of the $c$-axis after annealing, in agreement with Refs.~\onlinecite{kwei_structure_1989,maiser_pulsed_1998,tsukada_role_2005,matsumoto_synthesis_2009,matsumoto_generic_2009,krockenberger_emerging_2013}, and a small change of the in-plane lattice parameter in some samples.
$(00l)$ reflections in XRD patterns~\cite{supp} give $c_\text{SC}=\SIrange{12.080}{12.088}{\angstrom}$ for all superconducting films.
As established by extensive studies on electron-doped films fabricated by molecular-beam epitaxy~\cite{naito_intrinsic_1997,naito_epitaxy_2002,krockenberger_universal_2011} and pulsed laser deposition~\cite{gupta_deposition_1989,mao_deposition_1992,roberge_improving_2009,hoek_effect_2014}, the $c$-axis parameter can be used as a measure of the oxygen content. 
In these studies, the $c$ parameter is always larger than the optimal superconducting value $c_\text{SC}$, as we also observe in type B over-oxygenated samples, and decreases with the concurrent elimination of excess oxygen atoms during annealing.
Hence, a value $c<c_\text{SC}$ in type A samples indicates oxygen deficiency.

\section{Transport measurements}

% AS-GROWN / LOW TEMPERATURE ANNEALING

\begin{figure}
\includegraphics[width=1\columnwidth]{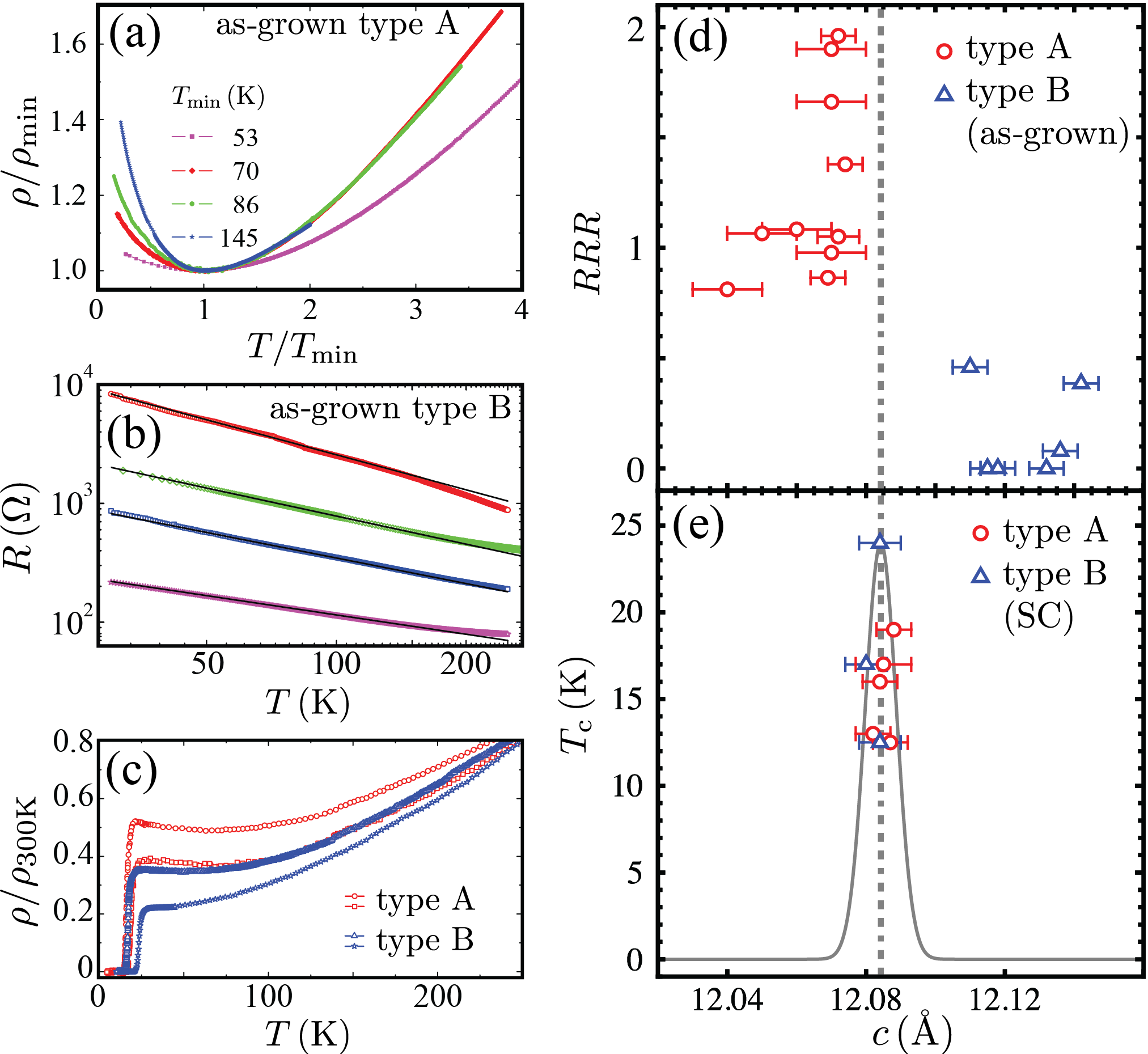}%
\caption{
(a) Resistivity as a function of temperature for type A samples, plotted on a log-log scale and normalized to the resistivity minimum.
(b) Resistance as a function of temperature for type B samples, plotted on a log-log scale.
Continuous lines $R(T) \propto T^{-\alpha}$ are the best fit to the data.
% with slopes $\alpha=0.54,0.72,0.79,0.98$. 
(c) Normalized resistivity as a function of temperature for type A and B samples after annealing, showing the superconducting transition at $T_\text{c}\lesssim\SI{24}{K}$.
(d) Residual resistivity ratio $RRR$ for as-grown samples (both types) and
(e) superconducting critical temperature $T_\text{c}$ of samples after annealing, as a function of the $c$-axis parameter.
The dashed gray line is the average value of $c_\text{SC}$ and the continuous smooth curve is a guide for the eye.
The annealing process was optimized for the sample reaching $T_\text{c}\approx \SI{24}{K}$. 
%In all samples, superconductivity emerges as soon as the structural reorganization is observed, regardless of the specific annealing conditions.
} 
\label{fig:RT}
\end{figure}

We measure the temperature dependence of the in-plane resistivity $\rho(T)$ with a four-probe method in the temperature range $\SIrange{1.6}{300}{\kelvin}$, before and after high-temperature annealing.
As-grown type A samples (fabricated in oxygen-free atmosphere) exhibit a crossover between metallic and insulating regimes identified by the resistivity minimum at temperature $T_\text{min}$, and a residual resistivity ratio $RRR=\rho(\SI{300}{K})/\rho(\SI{4.2}{K})>1$ [see \cref{fig:RT}(a)].
Furthermore, the resistivity exhibits a quadratic temperature dependence in the metallic region above $T_\text{min}$.
In electron-doped compounds, a quadratic resistivity dependence is usually found even above room temperature~\cite{armitage_progress_2010,bach_high_2011,sarkar_anomalous_2018,greene_strange_2020}.
As-grown type B samples (fabricated in mixed argon/oxygen atmosphere) exhibit instead a weak semiconductor-like temperature dependence of the resistivity with $RRR<1$ and $\rho(T) \propto R(T)\propto T^{-\alpha}$ [see \cref{fig:RT}(b)].

% HIGH TEMPERATURE ANNEALING

Annealing induces a modification of the oxygen content and a structural reorganization and redistribution of crystal defects and dislocations.
To disentangle these two effects, we performed high-temperature \emph{ex situ} annealing in oxygen-free, pure argon flow.
Despite different environmental growth conditions, structural, and electrical properties of type A and B samples, similar thermal treatments are needed to induce superconductivity.
All samples become superconducting after high-temperature annealing, regardless of the specific annealing conditions, and with similar critical temperatures $T_\text{c}\lesssim\SI{24}{K}$. 
In contrast, no superconducting transition and no structural change are detected after annealing at temperatures below \SI{900}{\celsius} and with the same environmental conditions, as reported elsewhere~\cite{guarino_correlation_2012}.

\Cref{fig:RT}(d-e) show the phase diagram of our NCCO samples as a function of the $c$-axis parameter, before and after high-temperature annealing, which is the main experimental result of this work.
In particular, \cref{fig:RT}(d) shows the residual resistivity ratio $RRR$ as a function of the $c$-axis parameter.
In the region $c<c_\text{SC}$ (sample type A), samples behave as weakly disordered metal with $RRR\gtrsim1$ and exhibit a metal-insulator crossover with minimum resistivity at $T_\text{min}$.
We observe $T_\text{min}$ up to $\SI{250}{K}$ and $RRR\approx\numrange{1}{2}$, with $RRR$ increasing with decreasing $T_\text{min}$.
In the region $c>c_\text{SC}$ (sample type B), samples behave as disordered systems with $RRR\lesssim1$, with a weakly semiconductor-like temperature dependence on the resistivity.
Most importantly, \cref{fig:RT}(e) shows the superconducting critical temperatures $T_\text{c}$ as a function of the $c$-axis parameter.
All samples achieve superconductivity after high-temperature annealing, accompanied by a structural change:
The $c$-axis increases in type A samples and decreases in type B samples. 
The superconducting regime is restricted to the value $c_\text{SC}=\SI{12.08}{\angstrom}$. 
Hence, high-temperature annealing induces not only superconductivity, but also a concurrent and systematic increase (in type A samples) or decrease (in type B samples) of the $c$-axis parameter toward the optimal value $c_\text{SC}$. 
This strongly suggests that the superconducting phase transition is induced by a structural reorganization and redistribution of oxygen atoms within the CuO$_2$ layers, charge reservoir layers, and in apical positions.
Moreover, the correlation between the concentration of apical oxygens and the $c$-axis parameter clearly points to the crucial role and impact of apical oxygens on the electronic properties after annealing.

\section{DFT calculations}

% DFT CALCULATIONS

\begin{figure}
\includegraphics[width=1\columnwidth]{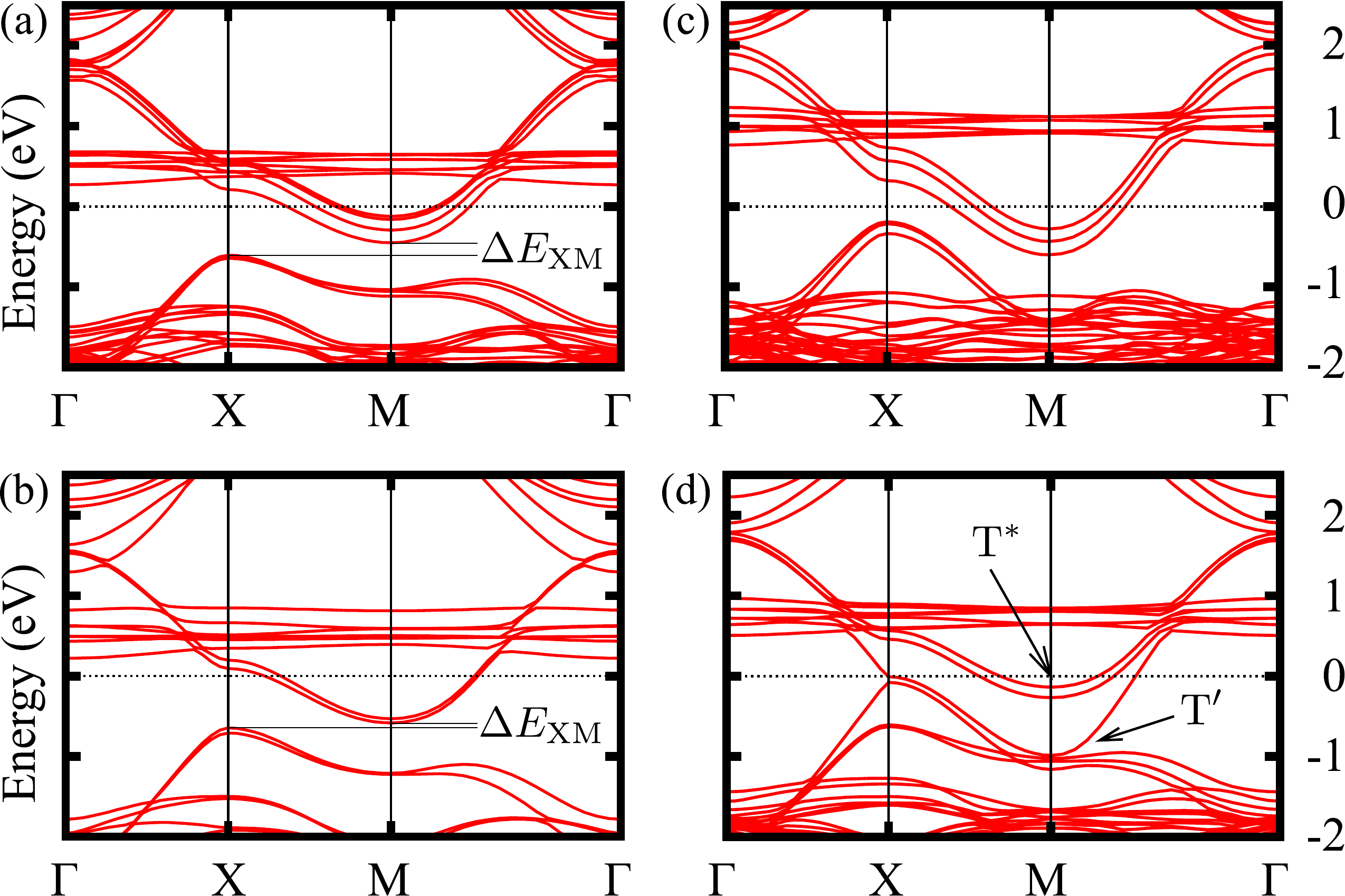}%
\caption{
DFT electronic band structure of NCCO for: 
(a) $\mathrm{T}^*$ structure with $x = 0.125$, 
(b) $\mathrm{T}^*$ structure with $x = 0.25$, 
(c) $\mathrm{T}'$ structure with $x=1/6\approx0.17$, and 
(d) $\mathrm{T}_\text{SC}=2\mathrm{T}^*+\mathrm{T}'$ structure with $x=1/6$, where we highlighted the dominant $\mathrm{T}'$ and $\mathrm{T}^*$ bands with larger contributions.
The Fermi level is set to zero.
The flat bands at $\SIrange{0.5}{1.5}{eV}$ above the Fermi level are the cerium 4$f$ bands. 
All other lower-energy bands are copper 3$d$ bands.
}
\label{fig:DFT}
\end{figure}

To understand the role of the structural reorganization of oxygen atoms, and the effects of the presence/absence of apical oxygens on the properties of the three types of samples, we modeled the system by DFT using the VASP~\cite{Kresse_96} package with plane-wave basis set and projector augmented wave method~\cite{Kresse_99}. 
In particular, as-grown type A samples are modeled by the $\mathrm{T}'$ structure as in \cref{fig:structure}, i.e., a crystal structure with no apical oxygens, according to its fabrication in oxygen-deficient atmosphere. 
As-grown type B samples are instead modeled by the $\mathrm{T}^*$ structure, i.e., a crystal structure with one apical oxygen for every copper atom, according to its fabrication in oxygen-rich atmosphere. 
Superconducting samples are modeled by a mixed $\mathrm{T}_\text{SC}=2\mathrm{T}^*+\mathrm{T}'$ structure, with two $\mathrm{T}^{*}$ cells and one $\mathrm{T}'$ cell alternating along the $c$-axis, i.e., a crystal structure with two apical oxygens for every three copper atoms.
This is justified by the experimental evidence that the $c$-axis parameter and, consequently, the number of apical oxygens, assume intermediate values between those measured for type A and B samples.
We performed DFT calculations by first relaxing the crystal structure to obtain the lattice parameters and compare them with the experimental ones and, in particular, with their characteristic hierarchical order.
We also computed the antiferromagnetic moments $m$ to monitor the intensity of antiferromagnetic correlations, which suppress superconductivity. 
(We study the $\mathrm{T}'$ and $\mathrm{T}^*$ structures of undoped \ce{Nd2CuO4} in the Appendix.)
The $\mathrm{T}^*$ structure shows a larger bandgap than the $\mathrm{T}'$ structure. 
The more correlated behavior of the $\mathrm{T}^*$ structure is confirmed by the magnetic moments, $m=0.38\mu_\text{B}$ and $m=0.43\mu_\text{B}$ for $\mathrm{T}'$ and $\mathrm{T}^*$ structures, respectively.
\Cref{fig:DFT}(a) and (b) show the results for the $\mathrm{T}^*$ structure for $x=0.125$ and $x=0.25$, respectively. 
In both cases and reasonably for all intermediate doping (including $x=0.17$), we have 
(i) an indirect bandgap $\Delta E_\text{XM}$ between the maximum of the lower Hubbard band at $\mathrm X$ (hole pocket) and the minimum of the upper Hubbard band at $\mathrm M$ (electron pocket),
(ii) strong antiferromagnetic correlations ($m=0.34\,\mu_\text{B}$ and $0.24\,\mu_\text{B}$, respectively, for $x=0.125$ and $0.25$), and
(iii) hole pockets away from the Fermi level. 
This strongly correlated scenario accounts for the insulating behavior of as-grown type B samples, in agreement with DMFT studies~\cite{weber_strength_2010}. 
\Cref{fig:DFT}(c) shows the results for the $\mathrm{T}'$ structure for $x=1/6\approx0.17$. 
The gap $\Delta E_\text{XM}$ closes, although antiferromagnetic correlations are still quite large ($m=0.27\,\mu_\text{B}$).
However, hole pockets are still far from the Fermi level. 
This scenario accounts for the (poor) metallic behavior of as-grown type A samples. 
Finally, \cref{fig:DFT}(d) shows the results for the $\mathrm{T}_\text{SC}$ structure for $x=1/6$. 
The bandgap $\Delta E_\text{XM}$ completely disappears, as well as antiferromagnetic correlations for the $\mathrm{T}'$ region ($m=0.04\,\mu_\text{B}$), while hole pockets are available right at the Fermi level (at the symmetry point $\mathrm{X}$) coexisting with electron pockets (at the symmetry point $\mathrm{M}$).
However, antiferromagnetic correlations in the $\mathrm{T}^*$ region are still relevant, being $m=0.36\,\mu_\text{B}$. 
The presence of holes at the Fermi level and the suppression of antiferromagnetic correlations indicate 
the emergence of hole superconductivity~\cite{dagan_hole_2007,li_hole_2019,hirsch_understanding_2019} in all samples after annealing.
Moreover, confirming the validity and accuracy of the chosen modelization and the consistency of the obtained results, the \emph{relaxed} values of the $a$ and $c$-axis parameters for $x=1/6$ are close to the experimental ones and, more importantly, in the same hierarchical order: 
For the $\mathrm{T}'$ structure (as-grown type A samples) $a=b=\SI{3.91}{\angstrom}$ and $c=\SI{12.01}{\angstrom}$, for the $\mathrm{T}^*$ structure (as-grown type B samples) $a=b=\SI{3.83}{\angstrom}$ and $c=\SI{12.26}{\angstrom}$, and for the $\mathrm{T}_\text{SC}$ structure (superconducting samples) $a=\SI{3.85}{\angstrom}$ and $c=\SI{12.18}{\angstrom}$. 
%The appearance of hole pockets can be understood as the effect of chemical pressure~\cite{ikeda_effects_2009} increased by the presence of apical oxygens?
The variation of the $c$-axis parameter can be understood in terms of \emph{level repulsion} between bands with dominant $\mathrm{T}'$ and $\mathrm{T}^*$ characters [see \cref{fig:DFT}(d)] that leads to a remodulation of the band structure, which definitely weakens antiferromagnetic correlations of the $\mathrm{T}'$ region and allows the emergence of holes right at the Fermi level (see also the Appendix).

\section{Conclusions}

% CONCLUSIONS

Concluding, the structural and the transport properties of NCCO samples shed new light on the microscopic mechanism underlying the annealing process, which is responsible for the onset of the superconductivity.
Indeed, our experiments indicate that the removal of the excess oxygen is not sufficient to trigger superconductivity:
Our oxygen-deficient samples, grown in oxygen-free atmosphere, become superconducting only after high-temperature annealing, which always occurs together with a change of the $c$-axis parameter.
This strongly indicates that the superconducting phase transition is induced by a microscopic structural reorganization, even in almost-optimally doped samples. 
Our theoretical analysis supports this conclusion, indicating a clear correlation between oxygen content, $c$-axis parameter, and superconductivity. 
In particular, no apical oxygens, or too many, stabilize strong antiferromagnetic correlations and keep holes away from the Fermi level. 
In contrast, an intermediate number of apical oxygens induces the suppression of antiferromagnetic correlations and allows to have holes available right at the Fermi level. 
Hence, the presence of a sizable number of apical oxygens is necessary to allow the emergence of hole superconductivity.
This points to the relevance of interlayer hoppings mediated by apical oxygens, analogously to the scenario evidenced in hole-doped cuprates~\cite{pavarini_band_2001,kim_apical_2018}.
Our experimental investigation and theoretical analysis provide strong evidence that the superconducting phase transition in electron-doped NCCO superconductors cannot be explained only in terms of changes of the oxygen content, but it necessarily requires a structural reorganization of the oxygen atoms in apical positions, which profoundly affects the electronic properties of the compound.

\begin{acknowledgments}
A.~G., A.~L., and G.~G. performed the experiments.
A.~G., A.~N., and P.~M. performed the experimental data analysis.
C.~A. and A.~A. performed the DFT calculations.
P.~M., A.~N., C.~A., and A.~A. wrote the manuscript.
A.~A. supervised the theoretical part.
A.~N. supervised the experimental part and the overall project.
All coauthors contributed to the scientific discussion and final version of the manuscript.
We thank M.~Wysokinski for useful discussions.
C.~A. is supported by the Foundation for Polish Science through the IRA Programme co-financed by EU within SG~OP\@.
C.~A. acknowledges the CINECA award under the IsC81 ``DISTANCE'' Grant for the availability of high-performance computing resources and support,
and the support of the Interdisciplinary Centre for Mathematical and Computational Modeling (ICM), University of Warsaw, under Grants No.~G73-23 and No.~G75-10\@. 
P.~M. is supported by the Japan Science and Technology Agency (JST) of the Ministry of Education, Culture, Sports, Science and Technology (MEXT), JST CREST Grant.~No.~JPMJCR19T, by the (MEXT)-Supported Program for the Strategic Research Foundation at Private Universities Topological Science, Grant~No.~S1511006, and by Japan Society for the Promotion of Science (JSPS) Grant-in-Aid for Early-Career Scientists, Grant~No.~20K14375.
A.~A. acknowledges support by MIUR under Project PRIN 2017RKWTMY.
\end{acknowledgments}

\appendix

\section{Sample preparation}

\ce{Nd_{2-x}Ce_xCuO4} (NCCO) films with a fixed cerium content in the range \numrange{0.16}{0.18} have been grown on $(100)$ \ce{SrTiO3} substrates by DC sputtering technique.
A single target of the stoichiometric \ce{Nd_{1.85}Ce_{0.15}CuO4} compound has been used as a sputtering source in an on-axis configuration with the substrate~\cite{guarino_fabrication_2013}. 
Type A and type B samples have been fabricated with a thickness varying in the range \SIrange{100}{200}{nm} at a total pressure of $\SI{1.7}{mbar}$ and heater temperature $\SI{850}{\celsius}$, respectively in pure argon and in mixed oxygen/argon atmosphere with ratio $\ce{O2}$/$\ce{Ar}$ in the range 2-14\%. 
A first \emph{in situ} annealing is performed at the same temperature. 
Type A films were \emph{in situ} annealed at different dwell times (20, 30, and 45 minutes) in the deposition chamber in vacuum at $10^{-5}\ \si{mbar}$ or $\SI{0.7}{mbar}$. 
Type B films were \emph{in situ} annealed at different dwell times (45 and 120 minutes) in the deposition chamber either in vacuum ($\SI{0.7}{mbar}$) or in argon atmosphere ($\SI{1.7}{mbar}$). 
A subsequent annealing was performed \emph{ex situ} in a Carbolite EHA 12/450B single-zone horizontal tube furnace with quartz/alumina tube and sealing flanges in flowing argon with 99.995\% purity, at a temperature set to $\SIrange{900}{950}{^\circ C}$, with a rate of $\SI{300}{^\circ C}$ per hour for both heating and cooling ramps, and with a dwell time of $\numrange{0.5}{2}$ hours, depending on the film thickness.
In all cases, the samples become superconducting after \emph{ex situ} annealing.
Despite different growing conditions and film thickness, the annealing temperature needed to induce superconductivity is the same for all films, while the annealing time depends on the film thickness.
The heating ramp used during annealing was optimized only for one sample, which therefore exhibits the nominal critical temperature $T_\text{c}\approx \SI{24}{K}$. 

\begin{figure}
\includegraphics[width=\columnwidth]{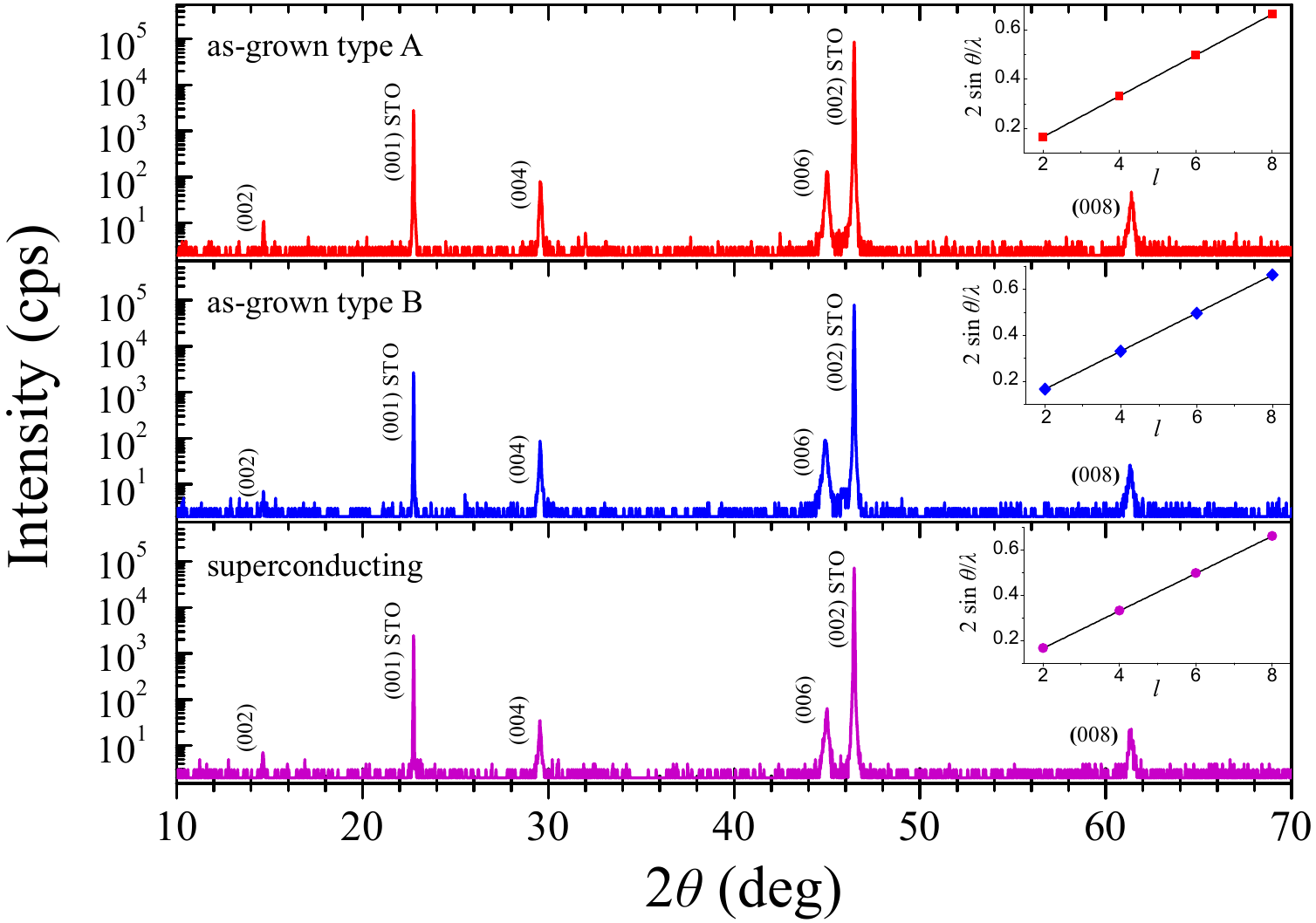}%
\caption{
X-ray diffraction patterns in units of counts per second for 
an as-grown type A sample, 
an as-grown type B sample, and 
a superconducting sample. 
The $(001)$ and $(002)$ peaks are due to the presence of the \ce{SrTiO3} substrate.
Insets show the linear fit of $2\sin\theta/\lambda$ as a function of the Miller index $l$.
The line slopes give the $c$-axis lattice parameter values. 
} 
\label{fig:1}
\end{figure}

\begin{figure*}
\includegraphics[width=1\textwidth]{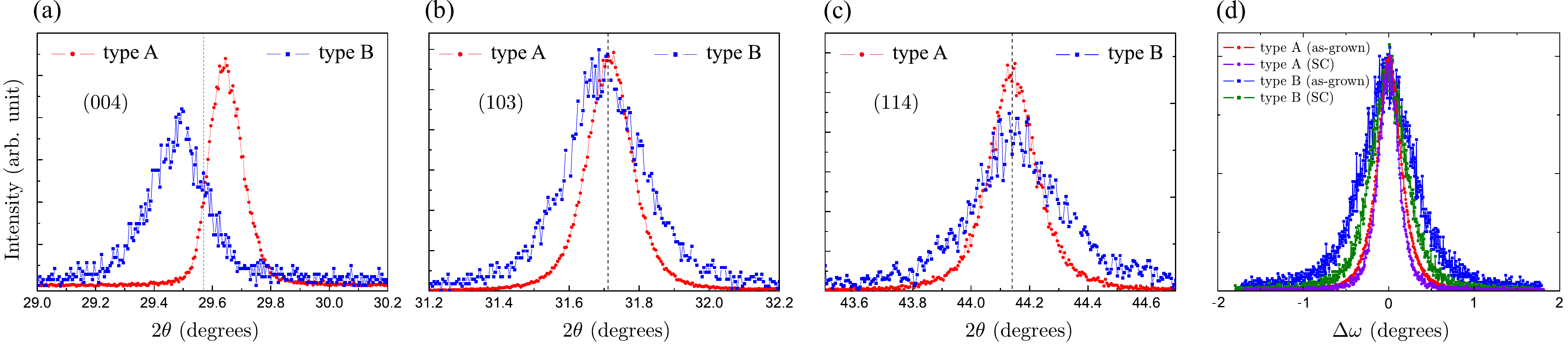}%
\caption{
Angular positions $2\theta$ of the (004), (103), and (114) peaks in panels (a), (b) and (c) respectively, for a type A (red circles) and a type B (blue squares) as-grown sample.
Dashed lines correspond to the $2\theta$ value measured in both samples after the annealing process.
(d) $\omega$-scans around the (004) reflections for a type A as-grown sample (red circles), a type A superconducting sample (purple circles), a type B as-grown sample (blue squares), and a type B superconducting sample (green squares), normalized to the maximum value.
} 
\label{fig:peaks}
\end{figure*}

Morphology, phase composition, and sample purity were inspected by scanning electron microscopy combined with wavelength-dispersive spectroscopy~\cite{guarino_transport_2011,guarino_fabrication_2013}, using an Oxford Scanning Electron Microscope Leo EVO 50 equipped with a wavelength-dispersive spectrometer.
Structural properties are obtained by high-resolution X-ray diffraction technique in a Philips X'Pert-MRD diffractometer equipped with a four circle cradle.

The electrical transport properties were investigated in a Cryogenic Ltd.\ cryogen-free cryostat equipped with an integrated cryogen-free variable-temperature insert operating in the range \SIrange{1.6}{300}{K}. 
In this system, the sample is cooled by a continuous helium gas flow and the temperature stability is within \SI{0.01}{K}. 
Sample temperature is measured via a LakeShore Temperature Controller model 350 connected to a LakeShore Cernox sensor. 
The electrical resistance measurements as a function of the temperature have been performed by a four-probe method, using a Keithley model 2430 as current source and a Keithley model 2182 as voltage meter. 
On selected films, in order to evaluate the resistivity, we realized microbridges with length $L=\SI{1}{mm}$, width $W=\SI{100}{\mu m}$ using a standard UV photolithography and wet etching in a $1\%$ solution of \ce{H3PO4} in pure water.
 
\section{X-ray diffraction measurements}

The structural properties of DC-sputtered NCCO films have been investigated by XRD technique. 
\Cref{fig:1} reports the typical $\theta$-$2\theta$ pattern of as-grown type A, as-grown type B, and superconducting samples.
Besides the substrate reflections, XRD patterns of as-grown, nonsuperconducting samples show only the four $(00l)$ diffraction peaks with $l=2,4,6,8$, which are characteristic of the $\mathrm{T}'$ tetragonal crystal structure, indicating a preferential growth with the $c$-axis perpendicular to the film surface and the absence of spurious phases.

The analysis of the $(00l)$ reflections allows to obtain directly the $c$-axis lattice parameter from the Bragg law, $2d\sin\theta=\lambda$, with $\lambda=\SI{1.54056}{\angstrom}$, $\theta$ the half of the angular peak position $2\theta$, and with $d=c/l$ in this case.
Insets of \cref{fig:1} report the quantity $2\sin\theta / \lambda$ as a function of the Miller index $l$ together with the linear best fit of the data. The fitting parameter gives a different $c$-axis parameter for each sample. 
In particular, the $c$-axis parameter measured in the superconducting sample is $c_\text{SC}=\SI{12.079\pm0.005}{\angstrom}$.
The value $c=\SI{12.069\pm0.005}{\angstrom}$ obtained for the type A film is shorter than $c_\text{SC}$, while the $c$-axis parameter $c=\SI{12.09\pm0.01}{\angstrom}$ of the ty pe B sample is longer. 
Hence, the most oxygenated type B samples behave as typical samples reported in previous studies, where the measured as-grown $c$-axis parameter is longer than the value measured in superconducting samples~\cite{kwei_structure_1989,maiser_pulsed_1998,tsukada_role_2005,matsumoto_synthesis_2009,matsumoto_generic_2009,krockenberger_emerging_2013}.
A value $c<c_\text{SC}$ is the peculiarity of our type A films. 

\Cref{fig:peaks} shows the angular positions $2\theta$ of the (004), (103), and (114) peaks respectively [(a) to (c)], for one type A and one type B as-grown samples compared with the value measured after annealing.
We found for the as-grown type A sample $c=\SI{12.04\pm0.01}{\angstrom}$ and $a,b=\SI{3.96\pm0.01}{\angstrom}$, while for the as-grown type B samples $c=\SI{12.11\pm0.01}{\angstrom}$, $a,b=\SI{3.94\pm0.02}{\angstrom}$. 

\Cref{fig:peaks}(d) shows the $\omega$-scans around the $(004)$ reflection for type A and B samples, as-grown and superconducting ($\omega$ is the x-ray incident angle). 
We observe a small difference between type A and B samples.
The full width at half maximum (FWHM) is in the range $\SIrange{0.5}{0.6}{^\circ}$ and $\SIrange{0.6}{0.7}{^\circ}$ respectively for as-grown type A and B samples, indicating a more uniform $c$-axis orientation for type A samples (i.e., better epitaxial growth). 
After annealing, we observe a slight reduction of the FWHM in all samples, which indicates a slight improvement in the mosaicity.

The decomposition products in NCCO thin films include \ce{NdCuO2}, \ce{Nd2O3}, and Nd-Ce-O phases~\cite{lin_parametric_1995}, which may in principle be present in samples annealed at high temperatures. 
In some of our samples, we observe a small fraction of the \ce{Nd_{1.85}Ce_{0.15}O3} phase in X-ray diffraction measurements.

\section{Computational details}

\begin{figure*}
\includegraphics[width=\textwidth]{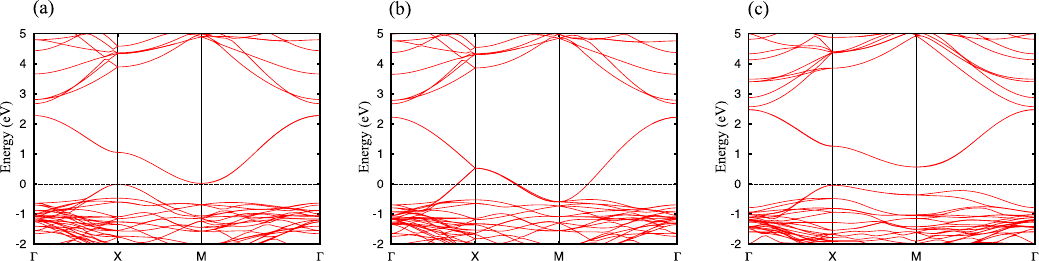}%
\caption{
Band structure of the Nd$_2$CuO$_4$.
(a)
$\mathrm{T}'$ phase with antiferromagnetic order,
(b)
$\mathrm{T}'$ phase with nonmagnetic atoms, 
(c)
$\mathrm{T}^{*}$ phase with antiferromagnetic order.
The value of the Coulomb repulsion is $U_\mathrm{Cu}= \SI{3.2}{eV}$. 
}
\label{DFT}
\end{figure*}

We have performed first-principles DFT calculations using the VASP~\cite{Kresse_96} package based on plane-wave basis set and projector augmented wave (PAW) method~\cite{Kresse_99}.
A plane-wave energy cut-off of \SI{450}{eV} has been used for the atomic relaxation and \SI{530}{eV} for the other calculations. 
A $k$-point grid of $8\times8\times2$ has been used for the atomic relaxation and $10\times10\times4$ for the other calculations. 
For the treatment of exchange correlation, Perdew-Burke-Ernzerhof~\cite{Perdew_08} generalized gradient approximation for solid has been considered, since it is accurate for the structural relaxation of the A$_2$BO$_4$ oxides bulk~\cite{Autieri_14} and other compounds with the transition-metal connected to 5 oxygen atoms~\cite{Asa_18,Asa_19}.

The analysis of the structural phases of compounds with 4$f$ electrons is a nontrivial problem in DFT due to the difficulties to catch the position of the energetic levels of the $f$ electrons~\cite{Kramer_19}.
Few works studied electron-doped cuprate superconductors using \emph{ab initio} techniques. 
Considering 4$f$ electrons in the core level, Bansil and coworkers were able to obtain the correct insulating groundstate for the undoped cases~\cite{Das_14,Kordyuk_18,Furness_18}.
We use the PAW with 3 frozen $f$ electrons for the Nd and without frozen $f$ electrons for the Ce. 
Using the PAW without frozen electrons for the Ce, the $\mathrm{T}'$ phase is always the ground state. 
Using the PAW with three frozen $f$ electrons for the Ce, we obtain the stabilization of the $\mathrm{T}^*$ phase, but this does not allow the Ce$^{+4}$ configuration experimentally observed.

We included the effects of the Hubbard $U$ on the Cu sites.
We scanned the values of $U_\mathrm{Cu}$ from $1$ to $\SI{4}{eV}$ for the undoped and used $J_\text{H}=0.15U$ for the Cu-3$d$ states, and we assumed the value of $U_\mathrm{Cu}=\SI{3.2}{eV}$ because for this value the undoped $\mathrm{T}'$ phase is a narrow gap semiconductor.
The Coulomb repulsion was applied also on the rare earth Nd and Ce (\SI{4}{eV}) and O (\SI{6}{eV}) but it is much less relevant since these electrons are far from the Fermi level.

To account for the G-type antimagnetism in \ce{Nd2CuO4} we use a $\sqrt{2}\times\sqrt{2}\times2$ supercell with 4 formula units. 
To investigate the structural properties as a function of doping, an additional calculation was done in the overdoped regime at $x=0.25$. 
Using Vegard's law, we estimated the lattice constants for $x=0.17$, which correspond to almost-optimally doping (as in our samples). 
Once we understood the structural properties, we study the electronic and magnetic properties of the compound relative to a value of the doping close to the experimental one. 
In order to do so, we used a $\sqrt{2}\times\sqrt{2}\times3$ supercell with 6 formula units. 
One Ce atom in 6 formula units will give the concentration of $x=1/6\approx0.17$.
This supercell can host 3 cuprate layers.
However, in order to reproduce the $\mathrm{T}^{*}$ phase we need an even number of layers. 
As a consequence, the $\sqrt{2}\times\sqrt{2}\times3$ supercell cannot host the $\mathrm{T}^{*}$ phase but it can host the $\mathrm{T}'$ phase and a mixed phase with two $\mathrm{T}^{*}$ cells and one $\mathrm{T}'$ cell alternating along the $c$-axis. 
In this article, we call this phase the $\mathrm{T}_\text{SC}=2\mathrm{T}^{*}$ + $\mathrm{T}'$ phase.

The most stable configuration of the Ce atoms is obtained when the Ce atoms are far from each other. 
This means that during growth the Ce atoms have a tendency to avoid each other, which points to a homogeneous distribution of these Ce atoms during the growth. 
In the most stable configuration of the $\mathrm{T}^*$ phase, the Ce atoms are not in the apical oxygen layer. 
In the most stable configuration of the mixed $\mathrm{T}_\text{SC}$ phase, the Ce atoms are closer to the \ce{CuO2} layers of the $\mathrm{T}'$ cell.

\section{DFT study of undoped N\lowercase{d}$_2$C\lowercase{u}O$_4$}

In this Section, we present the results of the undoped Nd$_2$CuO$_4$.
The $\mathrm{T}'$ and $\mathrm{T}^*$ phases of the Nd$_2$CuO$_4$ have the same stoichiometry but a different atomic position of the oxygen atoms and consequently of the atomic layers.
The $\mathrm{T}'$ phase consists of 4 atomic layers CuO$_2$/O/Nd$_2$/O while the $\mathrm{T}^{*}$ contains 3 atomic layers CuO$_2$/NdO/NdO. 
The different atomic composition of the atomic layers has a large influence on the lattice constant $c$ and consequently on the in-plane lattice constant too.
Considering just the effect of the packaging, we would expect that the $\mathrm{T}'$ phase with 4 atomic layers should have a larger $c$ lattice constant, but we also need to consider the effect of the charge. 
In an oversimplified ionic picture, the CuO$_2$ layers have a total charge $-2$, the O layers have a charge $-2$, the Nd$_2$ layers have a charge $+6$ while the NdO layers have a charge $+1$. 
Therefore, the 4 layers of the $\mathrm{T}'$ phase have charge $-2/-2/+6/-2$ while the 3 layers of the $\mathrm{T}^{*}$ phase have $-2/+1/+1$.
Due to the greater charge, the 4 layers of the $\mathrm{T}'$ phase attract each other much more than the 3 layers of the $\mathrm{T}^{*}$ phase resulting in a shorter $c$-axis of the $\mathrm{T}'$ phase. 
Therefore, there is an interplay and competition between the charge and the volume effect;
as a result, the $\mathrm{T}'$ phase has a shorter $c$-axis than the $\mathrm{T}^{*}$ phase. 
As a consequence of the shorter $c$, the $\mathrm{T}'$ phase presents a larger value of the in-plane lattice constant $a$. 
This simplified picture was verified in our DFT results.
We performed structural relaxation for the undoped case for the $\mathrm{T}'$ and $\mathrm{T}^{*}$ phases.
We obtained $a=\SI{3.91}{\angstrom}$ and $c=\SI{12.12}{\angstrom}$ for the $\mathrm{T}'$ phase: 
The total volume is $\SI{92.7}{\angstrom^3}$ per formula unit. 
We obtained $a=\SI{3.83}{\angstrom}$ and $c=\SI{12.34}{\angstrom}$ for the $\mathrm{T}^{*}$ phase:
The total volume is $\SI{90.3}{\angstrom^3}$ per formula unit.

The Cu states in $\mathrm{T}^{*}$ are more ionic due to the larger number of nearest-neighbor oxygen atoms, indeed the Cu $d$ orbitals are more localized and therefore the $\mathrm{T}^{*}$ phase is more insulating.
Instead, the $\mathrm{T}'$ phase is a semiconductor. 
Once we fixed the equilibrium atomic positions, we investigate the electronic properties scanning the value of $U_\mathrm{Cu}$.
We search for the critical value of $U_\mathrm{Cu}^{cr}$ such that the $\mathrm{T}'$ phase is insulating, we get the value $U_\mathrm{Cu}^{cr}=\SI{3.2}{ eV}$ for the $\mathrm{T}'$ phase and we assume this value for all the following calculations.
The band structure of the semiconducting $\mathrm{T}'$ phase is shown in \cref{DFT}(a).
We have the completely unoccupied upper Hubbard band between 0 and +\SI{2.2}{eV} above the Fermi level due to the $x^2-y^2$ orbital in the minority spin channel. The lower Hubbard band due to the $x^2-y^2$ orbital in the majority spin channel is completely occupied and entangled with other occupied Cu $d$ bands.
The band structure shows an indirect band gap with the maximum of the valence band at the X point and the minimum of the conduction band at the M point.
The gap in the DFT approach is opened by the interplay between $U_\mathrm{Cu}$ and the antiferromagnetic order: 
Indeed, the $\mathrm{T}'$ phase without magnetism shows a metallic phase with robust holes at the X point as shown in \cref{DFT}(b).
In the nonmagnetic phase, we also have a nonsymmorphic symmetry that produces a double degenerate band along the MX direction and a semi-Dirac point in X.
Performing the antiferromagnetic calculation for the $\mathrm{T}^{*}$ phase, we obtain the band structure shown in \cref{DFT}(c).
The band structure of the $\mathrm{T}^*$ phase shows a larger band gap and flatter Cu $d$ bands, but for the rest, we have the same properties as in the $\mathrm{T}'$ phase.

At $U_\mathrm{Cu}=\SI{3.2}{eV}$, the energy difference between the antiferromagnetic and the nonmagnetic phase is \SI{22}{meV} per formula unit for the $\mathrm{T}'$ phase and \SI{70}{meV} per formula unit for the $\mathrm{T}^{*}$ phase. 
Therefore, the $\mathrm{T}^*$ phase has a larger gap and its antiferromagnetic ground state is more robust. 
Increasing the value of $U$, the antiferromagnetic phase will become more stable but the scenario described here does not change qualitatively.

%\bibliography{bib}
%\bibliographystyle{prsty_no_etal_titles_doi_preprint_noemph}

\end{document}